\documentclass[preprint]{elsarticle}

\usepackage{lineno,hyperref,amsmath}
\modulolinenumbers[5]

\journal{Physics Letters B}

\usepackage{graphicx}
\usepackage{dcolumn}
\usepackage{bm}
\usepackage{hyperref}
\hypersetup{colorlinks=true, citecolor=blue, urlcolor=blue, linkcolor=blue}
\usepackage{color}
\usepackage{epsfig}
\usepackage{epstopdf}
\def\nuc#1#2{\relax\ifmmode{}^{#1}{\protect\text{#2}}\else${}^{#1}$#2\fi}
\begin{document}
  \newcommand {\nc} {\newcommand}
  \nc {\beq} {\begin{eqnarray}}
  \nc {\eeq} {\nonumber \end{eqnarray}}
  \nc {\eeqn}[1] {\label {#1} \end{eqnarray}}
  \nc {\eol} {\nonumber \\}
  \nc {\eoln}[1] {\label {#1} \\}
  \nc {\ve} [1] {\mbox{\boldmath $#1$}}
  \nc {\ves} [1] {\mbox{\boldmath ${\scriptstyle #1}$}}
  \nc {\mrm} [1] {\mathrm{#1}}
  \nc {\half} {\mbox{$\frac{1}{2}$}}
  \nc {\thal} {\mbox{$\frac{3}{2}$}}
  \nc {\fial} {\mbox{$\frac{5}{2}$}}
  \nc {\la} {\mbox{$\langle$}}
  \nc {\ra} {\mbox{$\rangle$}}
  \nc {\etal} {\emph{et al.}}
  \nc {\eq} [1] {(\ref{#1})}
  \nc {\Eq} [1] {Eq.~(\ref{#1})}
  \nc {\Ref} [1] {Ref.~\cite{#1}}
  \nc {\Refc} [2] {Refs.~\cite[#1]{#2}}
  \nc {\Sec} [1] {Sec.~\ref{#1}}
  \nc {\chap} [1] {Chapter~\ref{#1}}
  \nc {\anx} [1] {Appendix~\ref{#1}}
  \nc {\tbl} [1] {Table~\ref{#1}}
  \nc {\Fig} [1] {Fig.~\ref{#1}}
  \nc {\ex} [1] {$^{#1}$}
  \nc {\Sch} {Schr\"odinger }
  \nc {\flim} [2] {\mathop{\longrightarrow}\limits_{{#1}\rightarrow{#2}}}
  \nc {\IR} [1]{\textcolor{red}{#1}}
  \nc {\IB} [1]{\textcolor{blue}{#1}}
  \nc{\pderiv}[2]{\cfrac{\partial #1}{\partial #2}}
  \nc{\deriv}[2]{\cfrac{d#1}{d#2}}

\begin{frontmatter}

\title{Reliable extraction of the $dB({\rm E1})/dE$ for $^{11}$Be \\
from its breakup at 520~MeV/nucleon}

\author[ulb]{L.~Moschini\corref{cor1}}
\ead{laura.moschini@ulb.ac.be}
\author[ik,ulb]{P.~Capel}
\ead{pcapel@uni-mainz.de}

\cortext[cor1]{Corresponding author}
\address[ulb]{Physique Nucl\'eaire et Physique Quantique (C.P.~229)\\
Universit\'e libre de Bruxelles (ULB),
50 avenue F.D. Roosevelt, B-1050 Brussels, Belgium}
\address[ik]{Institut f\"ur Kernphysik, 
Johannes Gutenberg-Universit\"at Mainz,
Johann-Joachim-Becher Weg 45, 
D-55099 Mainz, Germany}

\begin{abstract}
We analyze the breakup of the one-neutron halo nucleus $^{11}$Be measured at 520~MeV/nucleon at GSI on Pb and C targets within an eikonal description of the reaction including a proper treatment of special relativity.
The Coulomb term of the projectile-target interaction is corrected at first order, while its nuclear part is described at the optical limit approximation.
Good agreement with the data is obtained using a description of $^{11}$Be, which fits the breakup data of RIKEN.
This solves the apparent discrepancy between the $dB({\rm E1})/dE$ estimations from GSI and RIKEN for this nucleus.
\end{abstract}


\end{frontmatter}



Since their discovery in the mid-80s halo nuclei have been the subject of intense experimental and theoretical studies \cite{Tan96,Rii13}.
These nuclei, located on the edge of the valley of stability exhibit a very exotic structure.
They are much larger than their isobars and this unusual size is qualitatively explained by a quantum-tunneling effect in which one or two loosely bound valence nucleons have a high probability of presence at a large distance from the other nucleons, far beyond the range of the nuclear interaction.
These nucleons hence form a diffuse halo surrounding a compact core.
The archetypes of halo nuclei are $^{11}$Be, with a one-neutron halo, and $^{11}$Li, with two neutrons in its halo.

Because of their short lifetime, these nuclei are mostly studied through reactions.
The breakup reaction, during which the halo nucleons dissociate from the core, is of special interest, as it reveals the internal structure of the projectile.
When the breakup is measured on a heavy target, like Pb, the reaction is dominated by the Coulomb interaction, and the dissociation is characterized by the E1 strength from the ground state to the core-halo continuum $dB({\rm E1})/dE$ \cite{AN13}.
In addition to its importance in the study of halo nuclei, this observable plays also a role in nuclear astrophysics, as it is related to the rate of radiative captures at low energy.
Coulomb breakup can thus provide an indirect method to infer cross sections of astrophysical interest \cite{BBR86,BHT03}.

Many measurements have been performed to constrain this value experimentally for various halo nuclei \cite{AN13}.
The Coulomb breakup of $^{11}$Be has been measured at 520~MeV/nucleon at GSI \cite{gsi_exp} and at 69~MeV/nucleon at RIKEN \cite{Fuk04}.
Surprisingly the E1 strengths inferred from both experiments differ significantly from one another.
A recent \emph{ab initio} calculation of $^{11}$Be provides a $dB({\rm E1})/dE$ in agreement with the RIKEN data \cite{calci}.
In this Letter, we reanalyze the GSI data to study the reason for this discrepancy.
We consider an eikonal description of the reaction \cite{Glauber,BC12}
with a correction of the Coulomb interaction \cite{bonaccorso-brink,CBS08}, which enables us to account for the Coulomb breakup and its interference with the contribution of the nuclear interaction.
We also use a proper treatment of special relativity \cite{satchler,winther-alder}, which seems to play a significant role at these energies \cite{bertulani_prl,ogata-bertulani}.
Following \Ref{CPH18}, we describe the structure of $^{11}$Be within a Halo-EFT, which has been fitted to the output of the \emph{ab initio} calculation of \Ref{calci}.
In addition to solve this longstanding issue, the model we develop in this work will provide a reliable tool to analyze similar measurements performed for both one- and two--nucleon halo nuclei at GSI and the recent RIBF facility at RIKEN.
It should therefore significantly contribute to the study of nuclear structure and astrophysics away from stability.

To describe the collision of a one-neutron halo nucleus on a target, we consider the following three-body model of reactions.
The projectile $P$ is described as a two-body quantal system with an inert core $c$, of mass $m_c$ and charge $Z_c e$, to which a neutron, of mass $m_n$, is loosely bound (the projectile is thus of mass $m_P=m_c+m_n$ and charge $Z_P=Z_c$).
Its internal Hamiltonian reads
\beq
H_0=-\frac{\hbar^2}{2\mu_{cn}}\Delta_{\ve{r}} + V_{cn}(\ve{r}),
\eeqn{e0}
where $\ve{r}$ is the $c$-$n$ relative coordinate, $\mu_{cn} = m_c m_n /m_P$ is their reduced mass, and $V_{cn}$ is an effective potential simulating their interaction.
In partial wave $l_j$, the eigenstates of $H_0$ read
\beq
H_0\ \phi_{ljm}(E,\ve{r}) = E\ \phi_{ljm}(E,\ve{r}),
\eeqn{e0a}
where $j$ is the total angular momentum resulting from the coupling of the orbital angular momentum $l$ with the spin of the halo neutron---the core is assumed spinless.
The eigenstates of $H_0$ of negative energy $E$ are discrete and correspond to the bound states of the projectile.
The positive-energy states describe the $c$-$n$ continuum, i.e. the broken-up projectile.

As mentioned before, we follow \Ref{CPH18} and consider a Halo-EFT description of $^{11}$Be.
In particular, we use the NLO $V_{cn}$ potentials developed in Sec.~V of \Ref{CPH18}.
These potentials are of Gaussian form and their depths in the $s_{1/2}$ and $p_{1/2}$ partial waves are adjusted to reproduce the long-range observables of the \emph{ab initio} bound-state wave functions: the one-neutron binding energy of the $\half^+$ ground state and the $\half^-$ excited states, respectively, as well as the corresponding asymptotic normalization constants (ANC).
They also provide $s_{1/2}$ and $p_{1/2}$ phaseshifts in agreement with the \emph{ab initio} prediction at low energy $E$.
In the other partial waves, $V_{cn}$ is set to zero.
As shown in \Ref{CPH18}, this description of $^{11}$Be leads to excellent agreements with the breakup cross sections measured at RIKEN on both Pb and C at about 70~MeV/nucleon \cite{Fuk04}.

The target $T$ is seen as a structureless body of mass $m_T$ and charge $Z_Te$, which interacts with the projectile constituents through optical potentials $V_{cT}$ and $V_{nT}$.
We denote by $\ve{R}$ the relative coordinate of the projectile center of mass to the target, setting the $Z$ axis along the incoming beam and calling $\ve{b}$ its transverse component.
Such a three-body model of the reaction provides a reliable framework to describe the elastic, inelastic and breakup channels of the collision \cite{BC12}.

Focusing on high beam energies, we naturally describe this collision within the eikonal approximation \cite{Glauber,BC12}, within which the three-body wave function behaves asymptotically after collision as
\beq
\Psi^{(m_0)}(\ve{R},\ve{r})\flim{Z}{+\infty} e^{iK_0Z} e^{i\chi(\ve{b},\ve{r})} \phi_0(E_0,\ve{r}),
\eeqn{e1}
where $\hbar K_0$ is the $P$-$T$ initial momentum, $\phi_0$ is the wave function of the projectile initial bound state of energy $E_0$ and total angular-momentum projection $m_0$, and $\chi$ is the eikonal phase that accounts for the interactions between the target and the projectile constituents
\beq
\chi(\ve{b},\ve{r})=-\frac{1}{\hbar v}\int_{-\infty}^\infty \left[V_{cT}(R_{cT})+V_{nT}(R_{nT})\right] dZ,
\eeqn{e2}
where $v=\hbar K_0/\mu$ is the $P$-$T$ relative velocity with $\mu=m_Pm_T/(m_P+m_T)$ the $P$-$T$ reduced mass.
The optical potentials $V_{cT}$ and $V_{nT}$ are assumed to be local and hence depend only on the distance between each of the constituents and the target $R_{cT}$ and $R_{nT}$.

This eikonal phase can be decomposed in its Coulomb and nuclear contributions
$\chi(\ve{b},\ve{r})= \chi^C_{PT}(b) + \chi^C(\ve{b},\ve{r}) + \chi^N(\ve{b},\ve{r})$.
The first term $\chi^C_{PT}(b) =  2 \eta \ln \left(K_0 b \right)$, with $\eta=Z_PZ_Te^2/4\pi\epsilon_0\hbar v$, describes the deflection of the projectile as a whole by the Coulomb field of the target \cite{bertulani-danielewicz}.

The second term results from the difference between the actual $c$-$T$ Coulomb interaction and the global $P$-$T$ Coulomb potential 
\beq
\chi^{C}(\ve{b},\ve{r}) = - \eta \int_{-\infty}^{\infty}\left( \frac{1}{|\ve{R}-\frac{m_n}{m_P}\ve{r}|} -\frac{1}{R} \right) dZ.
\eeqn{e5}
It describes the contribution of the Coulomb interaction to the tidal force that leads to the breakup of the projectile.
Because it decreases as $1/b$, its contribution to the breakup cross section diverges.
The problem is related to the adiabatic treatment of the $P$-$T$ interaction included in the usual eikonal description of reactions \cite{bonaccorso-brink,CBS08}.
In \Ref{bonaccorso-brink}, Margueron \etal\ have suggested to substitute the first-order term of the Coulomb phase $e^{i\chi^C}$ by the breakup amplitude at the first-order of the perturbation theory
$e^{i\chi^C}\rightarrow e^{i\chi^C} - i\chi^C + i\chi^{FO}$, where
\beq
\chi^{FO}(\ve{b},\ve{r}) &=& - \eta \int_{-\infty}^{\infty} e^{i \omega Z/v} \left( \frac{1}{|\ve{R}-\frac{m_n}{m_P}\ve{r}|} -\frac{1}{R} \right) dZ \label{e7a} \\
&=&-\eta\frac{m_n}{m_P}\frac{2\omega}{v}\left[K_1\left(\frac{\omega b}{v}\right)\frac{\ve{b}\cdot\ve{r}}{b} + i K_0\left(\frac{\omega b}{v}\right)z\right], 
\eeqn{e7b}
with $\hbar\omega=E-E_0$, the projectile excitation energy.
This correction has been shown to be very efficient compared to fully dynamical reaction models \cite{CBS08}.
Not only does it solve the aforementioned divergence issue, but it also restores some dynamical effects, such as postacceleration of the projectile fragments after the breakup, that are missing in the usual eikonal approximation.

The third term in the eikonal phase corresponds to the nuclear $c$-$T$ and $n$-$T$ interactions, which are usually described by optical potentials.
At the energies considered here it is difficult to find appropriate potentials, expecially for radioactive nuclei. Therefore, following \Ref{31Ne}, we rely on the optical limit approximation (OLA) of the Glauber theory \cite{Glauber,bertulani-danielewicz}.
In that approximation, the nuclear eikonal phase is obtained by averaging a profile function $\Gamma_{NN}$, which simulates the nucleon-nucleon interaction, over the nuclear density of the target $\rho_T$ and the projectile constituents $\rho_x$, where $x$ stands for $c$ or $n$
\beq
\chi^{OLA}_{xT}(\ve{b}_x)=  i \iint \rho_T(\ve{r'})  \rho_x(\ve{r''}) \Gamma_{NN}(\ve{b}_x-\ve{s'}+\ve{s''})d\ve{r'}\ d\ve{r''},~~~
\eeqn{e8}
where $\ve{b}_x$ is the transverse coordinate of $\ve{R}_{xT}$, $\ve{s'}$ and $\ve{s''}$ are the transverse components of the internal coordinate of the target ($\ve{r'}$) and $x$ ($\ve{r''}$), respectively.
In our three-body model of the reaction, the nuclear eikonal phase thus reads
$\chi^N(\ve{b},\ve{r}) = \chi^{OLA}_{cT}(\ve{b}_c) + \chi^{OLA}_{nT}(\ve{b}_n)$.
We consider the usual Gaussian form of the profile function and use the values of its parameters provided in \Ref{OLA_prof-func_param} for an energy of 550~MeV.
The densities used in \Eq{e8} for the $^{10}$Be core and the targets $^{208}$Pb and $^{12}$C are approximated by the two-parameter Fermi distributions of \Ref{OLA_densities_param}, in which the authors study a systematization of nuclear densities based on charge distributions extracted from electron-scattering experiments as well as on theoretical densities derived from Dirac-Hartree-Bogoliubov calculations.
For $\rho_n$, we consider a Dirac delta function.

The breakup cross section as a function of $c$-$n$ relative energy $E$ after dissociation, in the projectile center-of-mass restframe, reads \cite{GBC06,CBS08}
\beq
\lefteqn{\frac{d\sigma_{\rm bu}}{dE}= \sqrt{\frac{8\mu_{cn}}{\hbar^2E}}\frac{1}{2j_0+1}\sum_{m_0}\sum_{ljm}} \nonumber\\
\int_0^{\infty}&b db&\left|\left\langle\phi_{ljm}(E)\left| e^{i\chi^N}\left( e^{i\chi^C} -i\chi^C+i \chi^{FO}\right)\right|\phi_0(E_0)\right\rangle\right|^2.~~~~~
\eeqn{e12}

Since the reactions on which we focus have been measured at high energy, a proper treatment of special relativity must be considered.
To account for the relativistic kinematics, we take a leaf out of the book of Pang \cite{pang} and follow Satchler, who derives an eikonal approximation of the solution of the Klein-Gordon equation expressed in the $P$-$T$ CM frame \cite{satchler}.
He obtains solutions identical to those expressed above
but replacing the $P$-$T$ reduced mass $\mu$ by the reduced energy
\beq
\epsilon/c^2=M_P M_T/(M_P+M_T),
\eeqn{e18}
obtained from the relativistic masses $M_P=\gamma_Pm_P$ and $M_T=\gamma_Tm_T$ using
\beq
\gamma_i = \frac{x_i+\gamma_L}{\sqrt{1+x_i^2+2x_i\gamma_L}},
\eeqn{e13}
where $i=P$ or $T$, $x_P=m_P/m_T=x_T^{-1}$, and $\gamma_L = 1+E_{\rm Lab}/m_P c^2$, with $E_{\rm Lab}$, 
the projectile kinetic energy in the laboratory, and $c$ the speed of light \cite{satchler}.
We also use the relativistic $P$-$T$ relative momentum
\beq
\hbar K_0 = \frac{v \epsilon}{c^2},
\eeqn{e14}
which now depends on the relativistic $P$-$T$ relative velocity
\beq
\frac{v}{c} = \frac{m_P}{\epsilon} \sqrt{\gamma_P^2 -1}.
\eeqn{e15}
Also the Sommerfeld parameter $\eta$ depends on the relativistic velocity $v$.
Note that this kinematics choice is consistent with the way the OLA is implemented \cite{OLA_prof-func_param}.

In addition to this relativistic kinematics, we need to account for the fact that the matrix elements in the expression of the breakup cross section \eq{e12} have to be evaluated in the projectile center-of-mass restframe and that we have to make sure that the equations we use to describe the collision are Lorentz invariant.
This will be true if the potentials that simulate the interaction between the projectile constituents and the target transform as the time-like component of a four-vector:  $V_{PT}(\ve{b},Z,\ve{r}) \rightarrow \gamma V_{PT}(\ve{b},\gamma Z,\ve{r})$, where $\gamma = \left( 1-v_P^2/c^2 \right)^{-1/2}$ with $v_P/c = \sqrt{\gamma_P^2-1}/\gamma_P$ the $P$ velocity in the $P$-$T$ CM frame.
This transformation is well established for electromagnetic fields \cite{winther-alder}, while it is just a conjecture for the nuclear interaction \cite{bertulani_prl,ogata-bertulani}.
At the usual eikonal approximation, i.e.\ including the adiabatic approximation, all the phases ($\chi^C_{PT}$, $\chi^C$, and $\chi^N$), being integrated over $Z$, are not affected by that boost.
In these phases, the only influence of the relativistic correction is to replace the $P$-$T$ velocity $v$ by its relativistic expression \eq{e15}.
However, this is not true for the correction of the Coulomb phase at the first-order of the perturbation theory $\chi^{FO}$ \eq{e7a}, where the transformation brings a $\gamma$ factor multiplying the relativistic $v$ in the phase
\beq
\chi^{FO}(\ve{b},\ve{r})=-\eta\frac{m_n}{m_P}\frac{2\omega}{\gamma v}
\left[K_1\left(\frac{\omega b}{\gamma v}\right)\frac{\ve{b}\cdot\ve{r}}{b} + i K_0\left(\frac{\omega b}{\gamma v}\right)z\right],
\eeqn{e14b}
which is consistent with Winther and Alder's relativistic Coulomb excitation result \cite{winther-alder}.
\noindent In Tab.\ \ref{tab1} we list the values of the relativistic quantities $\gamma$, $\gamma_P$, $v/c$, $v_P/c$, $K_0$, and $\eta$, 
for the two reactions measured at GSI on C and Pb targets.
\begin{table}[h!]
\centering
\begin{tabular}{c|cccccc}
target & $\gamma$ & $\gamma_P$ &  $v/c$ & $v_P/c$ & $K_0$ (fm$^{-1}$) & $\eta$  \\
\hline
Pb & 1.4874 & 1.4872 &0.798  & 0.740 & 57.7 & 3.00 \\
C & 1.14090 & 1.14089 &0.931 & 0.481 & 28.8 & 0.188
\end{tabular}
\caption{Value of the relativistic quantities used for the two reactions measured at GSI on C and Pb targets.}
\label{tab1}
\end{table}

\begin{figure}
\includegraphics[width=\columnwidth]{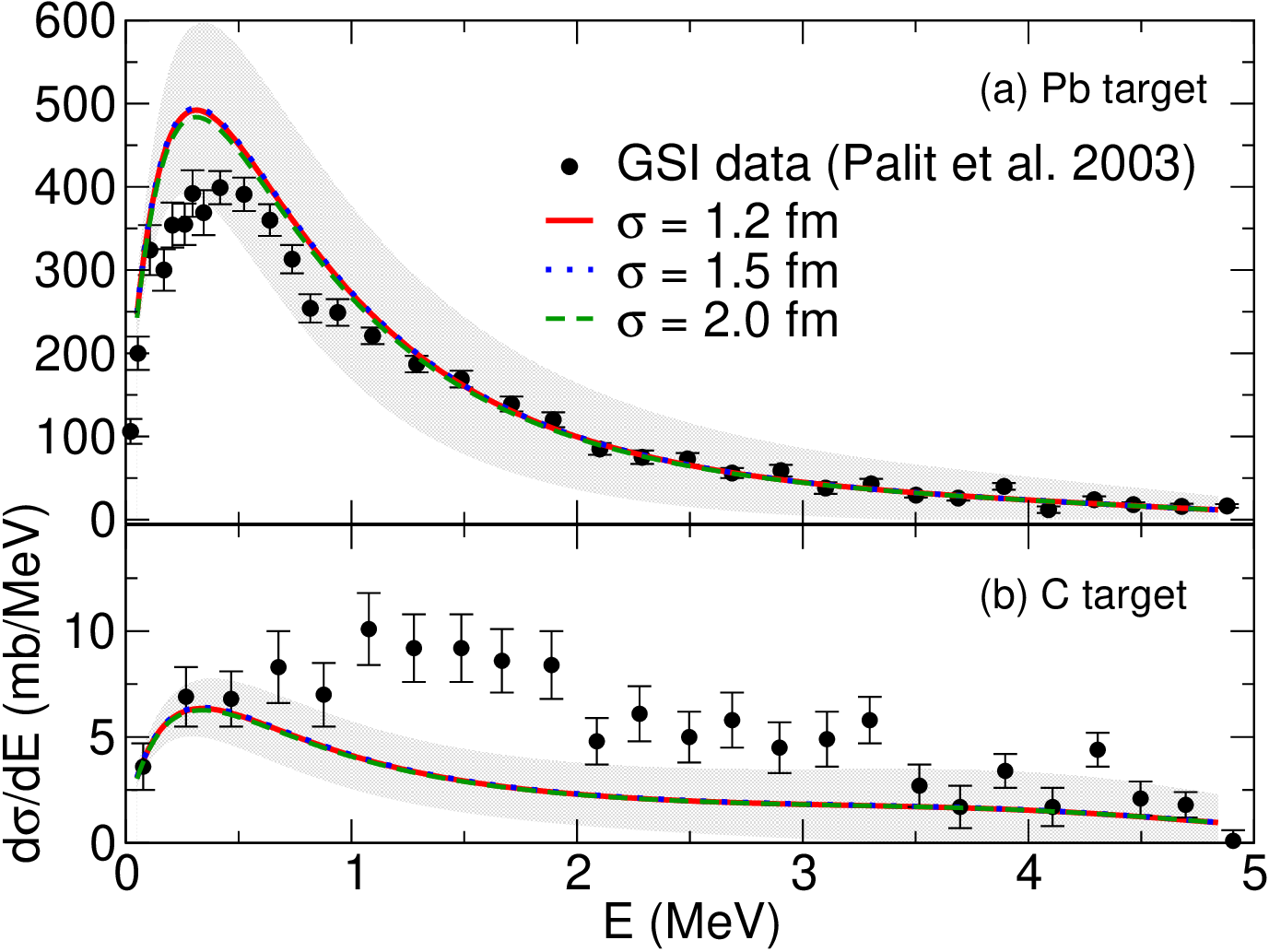}
\caption{\label{f2} Breakup cross section of $^{11}$Be on (a) $^{208}$Pb and (b)  $^{12}$C targets at 520~MeV/nucleon plotted as a function of relative energy $E$ between the $^{10}$Be core and the neutron after dissociation.
Results obtained with different Halo-EFT $c$-$n$ interactions are shown, as well as the NLO uncertainty band.
For comparison with the data of \Ref{gsi_exp}, the theoretical predictions are folded with the experimental energy resolution.}
\end{figure}

In \Fig{f2}, we plot the cross sections \eq{e12} obtained with this eikonal model of reactions for the breakup of $^{11}$Be on (a) Pb and (b) C at 520~MeV/nucleon, which correspond to the experiment performed at GSI \cite{gsi_exp}.
Let us first note the general agreement of our calculations with the data.
On Pb, our predictions fall nearly on top of the data and most of the experimental points are within the uncertainty band displayed as a gray area.
This band estimates the error made in truncating the Halo-EFT expansion of the $^{11}$Be description at NLO \cite{HJP17,CPH18}, 
it provides the order of magnitude of the uncertainty related to the missing degrees of freedom in the expansion, like the $d$-wave phase shifts or the core excitation.
The only disagreement is observed at very low energy, where the experimental uncertainty is the largest.
On C, we obtain results reminiscent to what has been obtained for the RIKEN data \cite{CPH18}: this NLO description of $^{11}$Be reproduces the general trend of the cross section, but misses a significant breakup strength in the 1--2~MeV energy range, which is due to the effect of the $\fial^+$ resonance, not included in this Halo-EFT model of the projectile.
This result confirms the need to include this degree of freedom within the description of $^{11}$Be to properly model the breakup on light targets \cite{Fuk04,CGB04}.

The calculations presented in \Fig{f2} have been performed with the three descriptions of $^{11}$Be developed in Sec.~V of \Ref{CPH18} using Gaussian potentials of widths $\sigma=1.2$~fm (solid red line), 1.5~fm (blue dotted line), and 2~fm (green dashed line).
These potentials provide bound-state wave functions with identical asymptotic behaviors, but which differ significantly in their interior.
Since all three potentials lead to identical results, we can conclude that high-energy breakup reactions of one-neutron halo nuclei are purely peripheral, as already observed at intermediate and low energy \cite{CN07}.
These reactions hence probe only the tail of the projectile wave functions, i.e. its ground-state ANC and the phaseshifts in the continuum.
Since these structure properties have been fitted onto the output of \emph{ab initio} calculations, the nice agreement with the data obtained here confirm the quality of the predictions of \Ref{calci}.

\begin{figure}
\includegraphics[width=\columnwidth]{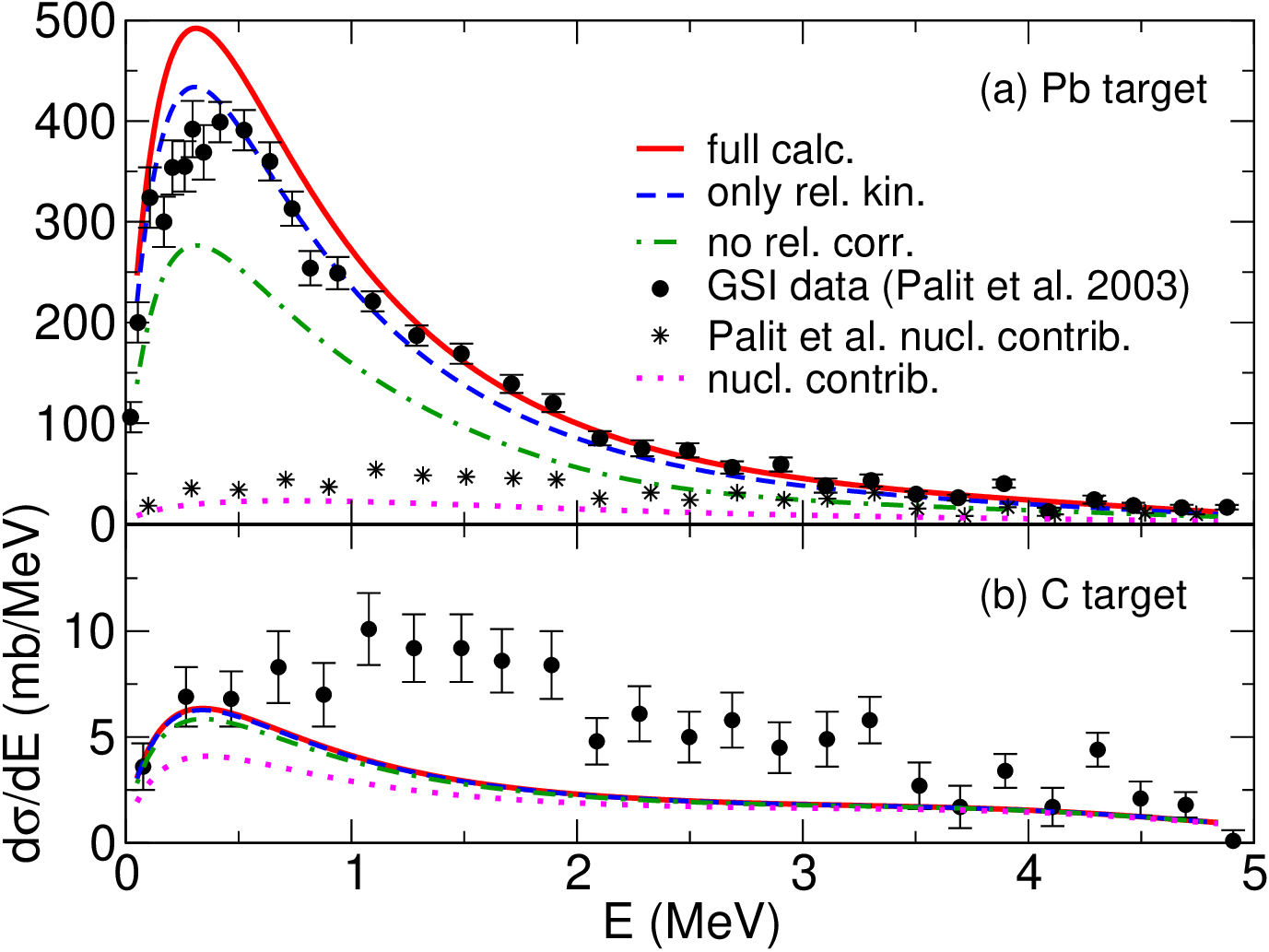}
\caption{\label{f3} Analysis of the relativistic corrections and nunclear contribution on the breakup cross section of $^{11}$Be on (a) $^{208}$Pb and (b)  $^{12}$C at 520~MeV/nucleon.}
\end{figure}
To estimate the role played by the nuclear interaction and the relativistic corrections in these reactions, we compare in \Fig{f3} our full calculation (red solid line) with the breakup cross sections obtained with only the nuclear term of the eikonal phase in our relativistic model of breakup (magenta dotted lines), with no relativistic correction (green dash-dotted line), and with only the relativistic kinematics (blue dashed line).
As expected, the reaction on Pb is strongly Coulomb dominated with a nuclear contribution of at most 5\%.
Interestingly, this contribution fits its GSI estimate [stars in \Fig{f3}(a)].
On C the reaction is strongly dominated by the nuclear interaction, although the Coulomb contribution remains non-negligible, especially at low energy.

The significance of the relativistic corrections is directly related to the dominance of the Coulomb interaction within the breakup process.
This can be quantitatively understood from the expression of the eikonal phases.
While the Coulomb phases  \eq{e5} and \eq{e14b} depend explicitly on $K_0$ and hence vary with the choice of the kinematics, the nuclear phase at the OLA  \eq{e8} depends only indirectly on this parameter through the projectile energy, and this dependence is rather weak at the beam energy considered here.
To properly analyze the Coulomb breakup of halo nuclei at high beam energy, a relativistic description of the reaction is thus needed \cite{bertulani_prl,ogata-bertulani}.
In that case, the kinematics is the dominant relativistic effect, although the boost required to compute the breakup matrix elements within the projectile center-of-mass restframe should not be overlooked.

The results presented in Figs.~\ref{f2} and \ref{f3} exhibit very little dependence on the parameters of the OLA used to simulate the $P$-$T$ nuclear interaction.
Our breakup cross sections barely change when other nuclear densities are considered \cite{1Bertulani_densities_param, 2Bertulani_densities_param} or when $\Gamma_{NN}$ is computed using the parameters tabulated at 425~MeV and 650~MeV \cite{OLA_prof-func_param}.
Our findings are thus robust.

\begin{figure}
\includegraphics[width=\columnwidth]{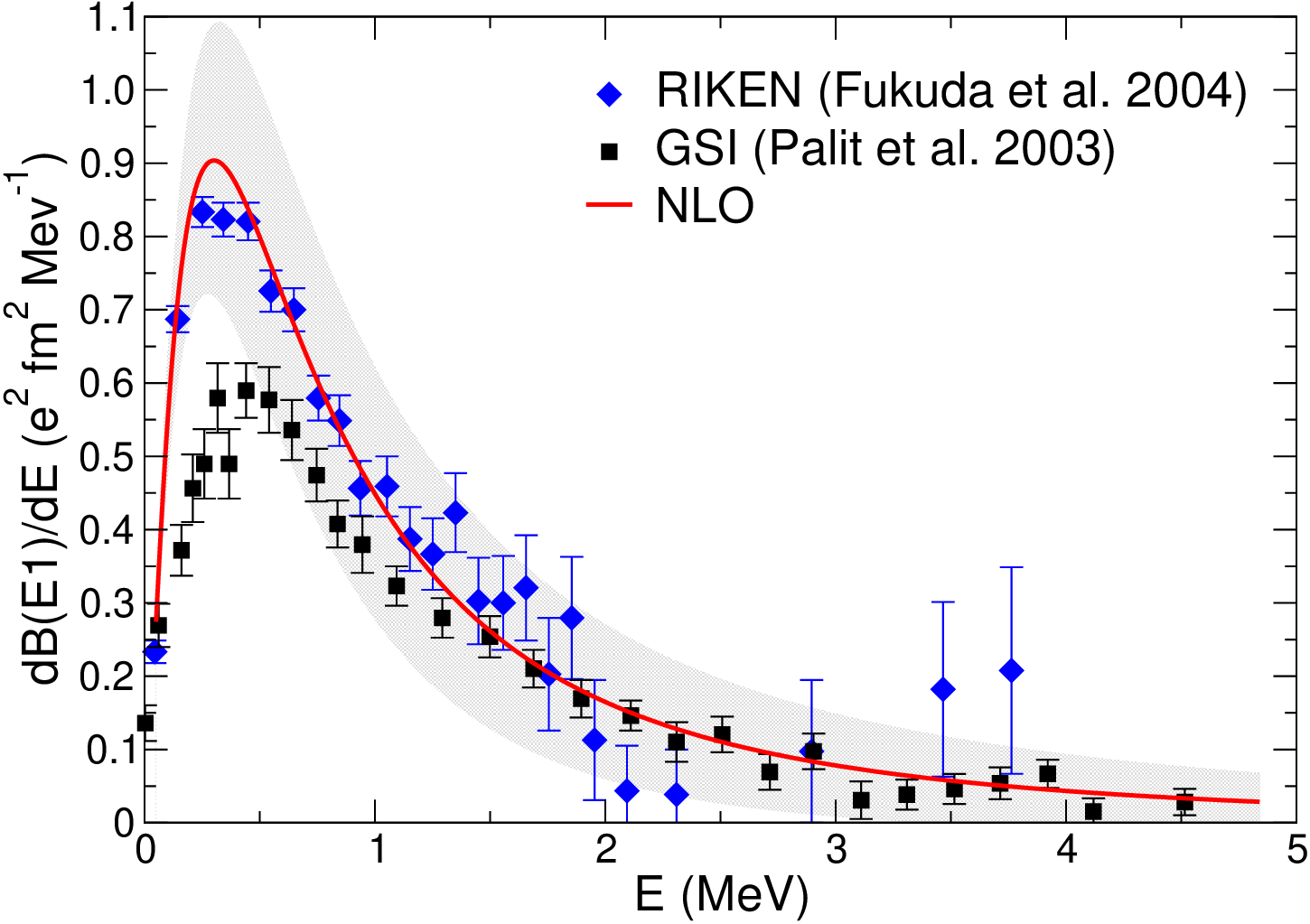}
\caption{\label{f4} The E1 strength computed from the Halo-EFT description of $^{11}$Be used in our reaction calculations ($\sigma=1.2$~fm) and the corresponding NLO uncertainty band compared to the values inferred at RIKEN \cite{Fuk04} and GSI \cite{gsi_exp}.}
\end{figure}

These nice results show that a good agreement with both most accurate measurements of the breakup of $^{11}$Be on Pb and C \cite{gsi_exp,Fuk04} can be obtained using one description of the projectile, which has been fitted onto the output of an \emph{ab initio} calculation of $^{11}$Be \cite{calci}.
Besides confirming the quality of the prediction of Calci \etal\ this suggests that this Halo-EFT description of $^{11}$Be should provide an accurate estimate of the E1 strength from the ground state to the continuum of this nucleus, which dominates the breakup mechanism on Pb.
This observable is plotted in \Fig{f4}.
Our NLO prediction is in excellent agreement with the results of Calci \etal\ (see Fig.~5 of \Ref{calci}) and with the values extracted from the RIKEN data \cite{Fuk04}.
However, it differs from the GSI ones,
even though both experiments are sensitive to the same projectile structure outputs: they are both peripheral and at these beam energies breakup cross sections on Pb are strongly dominated by the E1 strength.

Since both cross sections can be equally well reproduced using one model of $^{11}$Be, and hence one E1 strength, the discrepancy between the experimental estimates of the $dB({\rm E1})/dE$ for $^{11}$Be is most likely due to differences in the analysis of the data.
The analysis of the GSI experiment includes a proper treatment of relativity \cite{gsi_exp}, hence this cannot explain the difference observed in \Fig{f4}.
However, unlike the RIKEN experiment, they evaluate the Coulomb contribution---and thus the E1 strength---by subtracting an estimate of the nuclear contribution from the total breakup cross section.
The latter is extrapolated from the breakup cross section measured on C and,
as shown in \Fig{f3}, it is quite good.
However, this way of doing neglects the quantal interferences between the Coulomb and nuclear contributions to the dissociation.
As already observed by Typel and Shyam, both contributions do not really add up, especially at low energy (see Figs.~1 and 2 of \Ref{TS01}).
A mere subtraction of the nuclear contribution will induce too low a Coulomb estimate and therefore too low an E1 strength as observed in \Fig{f4}.
The analysis of the RIKEN data is less sensitive to this issue because it focuses on a measurement at forward angle, where the nuclear contribution is negligibly small (see, e.g., Fig.~3 of \Ref{TS01}).

The good results obtained in the present study indicate that our model of reaction can reliably account for relativistic effects in reactions involving loosely-bound nuclei at high energy while including both the Coulomb and the nuclear interactions between the projectile and the target, as well as their interferences, at all orders.
It is therefore an ideal tool to analyse this kind of reactions.
In a near future, we will use it to study more quantitatively the dynamics of the reaction, and in particular the significance of the interferences between the Coulomb and nuclear contributions to the breakup to confirm the present analysis.
We also plan to perform similar calculations for other one-nucleon halo nuclei, whose breakup has been measured at GSI or at the new RIBF facility in RIKEN like $^{15}$C \cite{Dat03}, $^8$B \cite{Sme99}, or $^{31}$Ne \cite{Nak09}.
We will also study the extension of our idea to collisions involving two-neutron halo nuclei like $^6$He \cite{Aum99} or $^{11}$Li \cite{Zin97}.
Hopefully, these new developments will provide the whole nuclear-reaction community with a reliable model to analyze breakup measurements that are used to study nuclear structure away from stability and estimate radiative-capture rates at astrophysical energies.

\section*{Acknowledgments}
We thank A.~Moro for his constructive suggestions on this project and the interesting discussion on our results.
This project has received funding from the European Union’s Horizon 2020 research and innovation programme under grant agreement No 654002.
It was also supported by the Deutsche Forschungsgemeinschaft within the Collaborative Research Centers 1044 and 1245, and the PRISMA (Precision Physics, Fundamental Interactions and Structure of Matter) Cluster of Excellence.
P. C. acknowledges the support of the State of Rhineland-Palatinate.


\end{document}